\documentclass[12pt]{iopart}

\usepackage{iopams}
\expandafter\let\csname equation*\endcsname\relax
\expandafter\let\csname endequation*\endcsname\relax
\usepackage{amsmath}
\usepackage{graphicx}
\usepackage{amsfonts}
\usepackage{amsbsy}
\usepackage{amssymb}
\usepackage{color}

\begin{document}
 
\title{Rare events and scaling properties in field-induced anomalous dynamics}

\author{R. Burioni}
\address{Dipartimento di Fisica e Scienza della Terra, Universit\`a degli Studi di Parma and INFN, Gruppo Collegato di Parma, viale G.P.Usberti 7/A, 43126 Parma, Italy}
\author{G. Gradenigo}
\address{CEA/DSM-CNRS/URA 2306, CEA Saclay, F-91191 Gif-sur-Yvette Cedex, France}
\author{A. Sarracino}
\address{ISC-CNR and Dipartimento di Fisica, Universit\`a Sapienza, p.le A. Moro 2, 00185 Roma, Italy}
\author{A. Vezzani}
\address{Centro S3, CNR-Istituto di Nanoscienze, Via Campi 213A, 41125 Modena, Italy and Dipartimento di Fisica e Scienza della Terra, Universit\`a degli Studi di Parma, viale G.P.Usberti 7/A, 43126 Parma, Italy}
\author{A. Vulpiani}
\address{Dipartimento di Fisica, Universit\`a Sapienza and ISC-CNR, p.le A. Moro 2, 00185 Roma, Italy}

\ead{raffaella.burioni@fis.unipr.it,ggradenigo@gmail.com, \\ alessandro.sarracino@roma1.infn.it, alessandro.vezzani@fis.unipr.it\\
angelo.vulpiani@roma1.infn.it}


\begin{abstract}
We show that, in a broad class of continuous time random walks (CTRW),
a small external field can turn diffusion from standard into
anomalous.  We illustrate our findings in a CTRW with trapping, a prototype of
subdiffusion in disordered and glassy materials, and in the L\'evy
walk process, which describes superdiffusion within inhomogeneous
media. For both models, in the presence of an external field, rare
events induce a singular behavior in the originally Gaussian
displacements distribution, giving rise to power-law
tails. Remarkably, in the subdiffusive CTRW, the combined effect of
highly fluctuating waiting times and of a drift yields a non-Gaussian
distribution characterized by long spatial tails and strong anomalous
superdiffusion.
\end{abstract}

\maketitle

\section{Introduction}

Large fluctuations and rare events play an important role in many
physical processes. Their contribution strongly influences physical
observables in laser cooling~\cite{bardou}, liposome
diffusion~\cite{wang}, heteropolymers~\cite{tang}, chaotic
systems~\cite{sanchez} and non-equilibrium
relaxation~\cite{bouchaud,ribiere}. In these systems, heterogeneous
spatial structures or temporal inhomogeneities can give rise to
\emph{anomalous} transport: the mean square displacement (MSD) of a
tagged particle is not linear in time, $\langle x^{2}(t)\rangle\sim
t^{2/\gamma}$, with $\gamma \neq 2$, and one observes either
subdiffusion, $\gamma>2$, or superdiffusion, $\gamma<2$.  However,
when the fluctuations in the microscopic dynamics are not too large,
i.e. the tails of their probability distributions decay fast enough,
anomalous transport is suppressed, and standard diffusion can occur
also in the presence of inhomogeneities. In these subtle situations,
detecting large deviations and establishing the heterogeneous nature
of the underlying microscopic dynamics can be difficult.  We show here
that the presence of an arbitrary small external field can induce an
anomalous growth of fluctuations even when the unperturbed behavior
remains Gaussian. This is an example, similar to others recently
pointed out in~\cite{beni1,beni3}, of how the response to an external
field may feature anomalous effects in the case of nonlinear
dynamics.  These topics can find application for instance
in probe-based microrheology, for the study of the motion of a tracer particle
in a disordered system when an external force is applied~\cite{wilson}.
We remark that, for our heterogeneous systems, the
Einstein relation established
in~\cite{barkai,jespersen,villamaina,bettolo,beni2} proves a
proportionality between $\langle x^{2}(t)\rangle$ and the drift
$\langle x(t)\rangle_\epsilon$, where $\langle\dots\rangle_\epsilon$
denotes the average over the process with an external field
$\epsilon$.

For standard diffusion governed by the Fick's law, the Gaussian form
of the probability density function (PDF) of displacements at time
$t$, $P_0(x,t)$, is not altered by the presence of a field $\epsilon$:
\begin{equation}
P_{\epsilon}(x,t) \simeq  \frac{1}{\sqrt{4 \pi D t}}\exp\left[
  -\frac{(x-v_\epsilon t)^2}{4Dt}\right] ,
\label{gaussian}
\end{equation}
where $D$ is the diffusion coefficient. The field induces a finite
drift, $\langle x(t) \rangle_\epsilon /t\equiv v_\epsilon\not= 0$, and
a \emph{typical} displacement $l_T(t)$, defined as the maximum of the
PDF, growing as $l_T(t) = v_\epsilon t$.

When the microscopic dynamics is characterized by large space
inhomogeneities and/or several time scales, the action of a field can
induce relevant deviations from the Gaussian
behavior~\cite{bouchaud}. We study here the properties of
$P_\epsilon(x,t)$ in the presence of a field for models with scale
invariant distribution of trapping times and with scale invariant
distributions of displacements. We consider a continuous time random
walk (CTRW)~\cite{bouchaud} with trapping, where a Brownian particle is 
trapped for a time interval distributed according to a given PDF, which shows
subdiffusive dynamics and mimics the slow activated dynamics of
complex fluids~\cite{angelani}. It embeds the waiting time
distribution
\begin{equation}
  p_\alpha(\tau)\sim\tau^{-(\alpha+1)},
  \label{levy}
\end{equation}
where $\alpha>0$ is the exponent characterizing the slow decay of time
distribution.

Fat tails also occur in displacements distribution, and in particular
they characterize the L\'evy-like motion in heterogeneous
materials~\cite{levitz,havlin,Levyrandom,Levy2d,benichou}, or in
turbulent flow~\cite{shles1985}, where transport is realized through
increments of size $l$ with distribution $p_\alpha(l)\sim
l^{-(1+\alpha)}$~\cite{barthelemy,bertolotti,klages} and
superdiffusive dynamics can occur.  Interestingly, it has been
recently found~\cite{lechenault} that even in a nearly arrested
granular assembly the displacements of grains follow a L\'evy
distribution, producing an unexpected superdiffusive
dynamics. Moreover, as predicted in~\cite{bouchaud}, and found in
numerical simulations of complex liquids~\cite{winter,heuer}, a
superdiffusive dynamics may be induced by an external field also in
trap-like disordered systems.

For vanishing field, i.e. $\epsilon=0$, if the tails of the steps and waiting times
distributions decay slowly enough, both the CTRW with trapping and the L\'evy
walk lead to anomalous transport, subdiffusive and superdiffusive,
respectively. In these regimes, motion is largely influenced by rare
events and $P_0(x,t)$ is not Gaussian, presenting, 
up to distances of order $\ell(t)$, the generalized 
scaling form $P_0(x,t) \sim t^{-1/z} F(x/t^{1/z})$, 
where $\ell(t)\sim t^{1/z}$ is the scaling length. In the case of fast
asymptotic decay of $F(y)$, the behavior of the moments is given by
$\langle x^{n}(t) \rangle \sim \ell^n(t)$.  However, due to rare
events, also $F(y)$ can present a slow decay. In this case the scaling
is broken for $x\gg \ell(t)$ and the system can feature ``strong''
anomalous behavior, i.e. the moments of the distribution are not a
power of $\ell(t)$~\cite{castiglione}.  On the other hand, when the
fluctuations of $p_\alpha(\tau)$ or of $p_\alpha(l)$ are not too
large, $P_0(x,t)$ is Gaussian.

In this paper we show that, remarkably, the presence of a drift can
induce an anomalous non-Gaussian dynamic even for regimes where
equilibrium measurements show a standard diffusive behavior. By
solving the master equations of the two processes mentioned above
in the one-dimensional case, we
determine the PDF in the presence of an external field.  We evidence
that, as expected, in the region of anomalous transport, the
distributions are very sensitive to the presence of a drift. More
surprisingly, in the regimes where the form of the distribution is
simply Gaussian at equilibrium, we find that the field can induce a
non-Gaussian shape of $P_\epsilon(x,t)$.  In the CTRW with trapping, we
identify an interval of values of $\alpha$ where the combined effect
of highly fluctuating waiting times and drift gives rise to a
perturbed non-Gaussian PDF spreading out \emph{superdiffusively}. Our
main result is that, even when drift and diffusion are standard, an
underlying anomalous behavior can be singled out by studying the
response to external perturbations.  An arbitrary small field can
induce a transition from standard Gaussian diffusion to a
\emph{strong} anomalous one.

\section{Models and scaling hypotheses}

In the CTRW with trapping, a particle moves with probability $1/2$ from $x$ to
$x \pm \delta_0$, where $\delta_0$ is constant.  The main results do
not change with a symmetric distribution of $\delta_0$ with finite
variance. Between successive steps, the particle waits for a time
$\tau$ extracted from the distribution~(\ref{levy}).  In the L\'evy walk model
again there are time intervals of duration $\tau$ extracted from
the distribution~(\ref{levy}), but the particle, during each time lag, moves at a
constant velocity $v$, chosen from a symmetric distribution with
finite variance, and performs displacements $\tau v$. Here we consider
$v=\pm v_0$ with equal probability and $v_0$ constant. In both models we introduce
a lower cutoff $\tau_0=1$ in the distribution~(\ref{levy}) so that, taking into account 
of the normalization, we have $p_\alpha(\tau)=\alpha \Theta(\tau-1) \tau^{-(1+\alpha)}$,
where $\Theta(x)$ is the Heaviside step function.
The value of $\tau_0$ does not change the behavior of the model apart from a 
suitable rescaling of the constants.

In the CTRW with trapping, the external field is implemented by unbalancing
the jump probabilities, i.e. setting to $(1\pm\epsilon)/2$ the
probability of jumping to the right or to the left, respectively. For
the L\'evy walk, in~\cite{sokolov,gradenigo} it has been shown that
the natural way to drive the system out of equilibrium is to apply an
external field accelerating the particle during the walk, so that the
distance traveled after a scattering event is $\delta= \pm v_0 \tau +
\epsilon \tau^2$. In the following we will consider a positive bias $\epsilon>0$.

Inspired by the Gaussian case, the simplest generalization of
Eq.~(\ref{gaussian}) for $\epsilon\ne 0$ is:
\begin{equation} 
P_\epsilon(x,t) \sim t^{-1/z} F[(x-v_\epsilon t)/t^{1/z}].
\label{scaling}
\end{equation}
When $\epsilon=0$, the left-right symmetry implies that $v_\epsilon=0$
and $F(y)$ is an even function.  In systems with power-law
distributions, Eq.~(\ref{scaling}) is made slightly more complex by
the occurrence of field-induced \emph{rare events}: in L\'evy walks a
fat tail due to large displacements in the direction of the field must
be considered, whereas in the CTRW with trapping we find a distribution
uniformly shifting in time in the direction of the field, plus an
algebraic tail for small values of $x$, originated from particles with
a large trapping time.  In both situations the simple scaling form of
Eq.~(\ref{scaling}) leads to inconsistencies in the behavior of the
moments of arbitrary order $\langle x^n(t)\rangle_\epsilon $.  Indeed
there is a cut-off in the largest distance from the peak of the
distribution at which a particle can be found in both models.  In the
L\'evy walk at time $t$, there can be no displacement exceeding $v_0 t
+ \epsilon t^2$, so that for large times we have:
\begin{equation} 
P_\epsilon(x,t) \sim t^{-1/z} F[(x-v_\epsilon
  t)/t^{1/z}]~\Theta\left(\epsilon t^2-x\right).
\label{scaling2}
\end{equation}
In the CTRW
with trapping, the factor $\Theta(\epsilon t^2-x)$ in Eq.~(\ref{scaling2}) has to be
replaced by $\Theta(x)$, which cuts off the power-law tail due to
particles with large persistence time at the origin.  In
Eq.~(\ref{scaling2}) there are three characteristic lengths: the
length-scale of the peak displacement, $l_T(t) \sim t $; the
length-scale for the collapse of the function, $l(t)\sim t^{1/z}$; the
length-scale of the largest displacement from the peak of the
distribution, $l_{e}(t) \sim t^2$.

\section{Solution of the master equation for the CTRW with trapping}

The scaling arguments~(\ref{scaling2}) can be proved by writing
the master equations relating, at two subsequent scattering events,
$P_\epsilon(x,t)$ with the function $Q_\epsilon(x,t)$, i.e. the
probability of being scattered in $x$ at time $t$~\cite{klafter,schmi}.  For
the CTRW with trapping we have:
\begin{equation}   
Q_\epsilon(x,t)=\int_0^t \left [\frac{1 +\epsilon}{2}
  Q_\epsilon(x-\delta_0,t-t') 
+ \frac{1 -\epsilon}{2}
  Q_\epsilon(x+\delta_0,t-t')\right]p_\alpha(t') dt' +\delta(t)\delta(x) 
\label{me1}
\end{equation}
with
\begin{equation}   
P_\epsilon(x,t)=\int_0^t Q_\epsilon(x,t-t') \int_{t'}^\infty p_\alpha(\tau)  d\tau dt',
\label{me2}
\end{equation} 
where $\delta(x)$ is the Dirac delta. We solve Eq.~(\ref{me1}), taking
the time and space Fourier transform, $P_\epsilon(x,t)\to \tilde P_\epsilon(k,\omega)$
and $Q_\epsilon(x,t)\to \tilde Q_\epsilon(k,\omega)$. From Eq.~(\ref{me1}) 
we obtain
\begin{equation}   
\tilde Q_\epsilon(k,\omega)=\frac{1}{1-[\cos(k \delta_0)+i\epsilon \sin(k \delta_0)]\int p_\alpha(t')e^{i \omega t'} dt'}.
\label{me3}
\end{equation}
Then, taking the Fourier transform of  Eq.~(\ref{me2}) we get 
\begin{equation}   
\tilde P_\epsilon(k,\omega)=\tilde Q_\epsilon(k,\omega) \int e^{i \omega t'}\int_{t'}^\infty p_\alpha(\tau) d\tau dt. '
\label{me4}
\end{equation}
We then substitute expression Eq.~(\ref{me3}) into Eq.~(\ref{me4}) and finally consider 
the asymptotic regimes where simultaneously 
$k\to 0$ and $\omega\to 0$, neglecting subleading terms. 
 
For $\epsilon=0$, the results are well
known~\cite{klafter}:
\begin{equation}   
\tilde P_0(k,\omega)\approx
\begin{cases}
\frac{C_1 \omega^{\alpha-1}}{(\delta_0 k)^2/2+ C_2 \omega^{\alpha}}\qquad  \mathrm{if}\qquad 0<\alpha<1 \\
\frac{\langle t \rangle}{(\delta_0 k)^2/2+ i \langle t \rangle  \omega}\qquad  \mathrm{if}\qquad \alpha >1,\\
\end{cases}
\label{four_tr_ep0}
\end{equation}
where $\langle t^p \rangle =\int t^p p_\alpha(t) dt$.
The complex number $C_i$ depends on the sign of $\omega$:
\begin{equation} 
C_1 (\alpha)=\int_0^\infty \frac{\cos(t)}{t^\alpha}dt+  i \ {\rm sign}(\omega) \int_0^\infty \frac{\sin(t)}{t^\alpha}dt
\end{equation}  
and 
\begin{equation} 
C_2(\alpha)= i \ {\rm sign}(\omega)\ C_1(\alpha)=- \int_0^\infty \frac{\sin(t)}{t^\alpha}dt+  i \ {\rm sign}(\omega) \int_0^\infty \frac{\cos(t)}{t^\alpha} dt,
\end{equation} 
with $0<\alpha<1$. For $\alpha> 1$, Eq.~(\ref{four_tr_ep0})
corresponds to standard diffusion, while for $0< \alpha<1$ the
Gaussianity is broken and the dynamics is characterized by a
subdiffusive scaling length $\ell(t)\sim t^{\alpha/2}$.

\begin{figure}[!t]
\includegraphics[width=.7\columnwidth,clip=true]{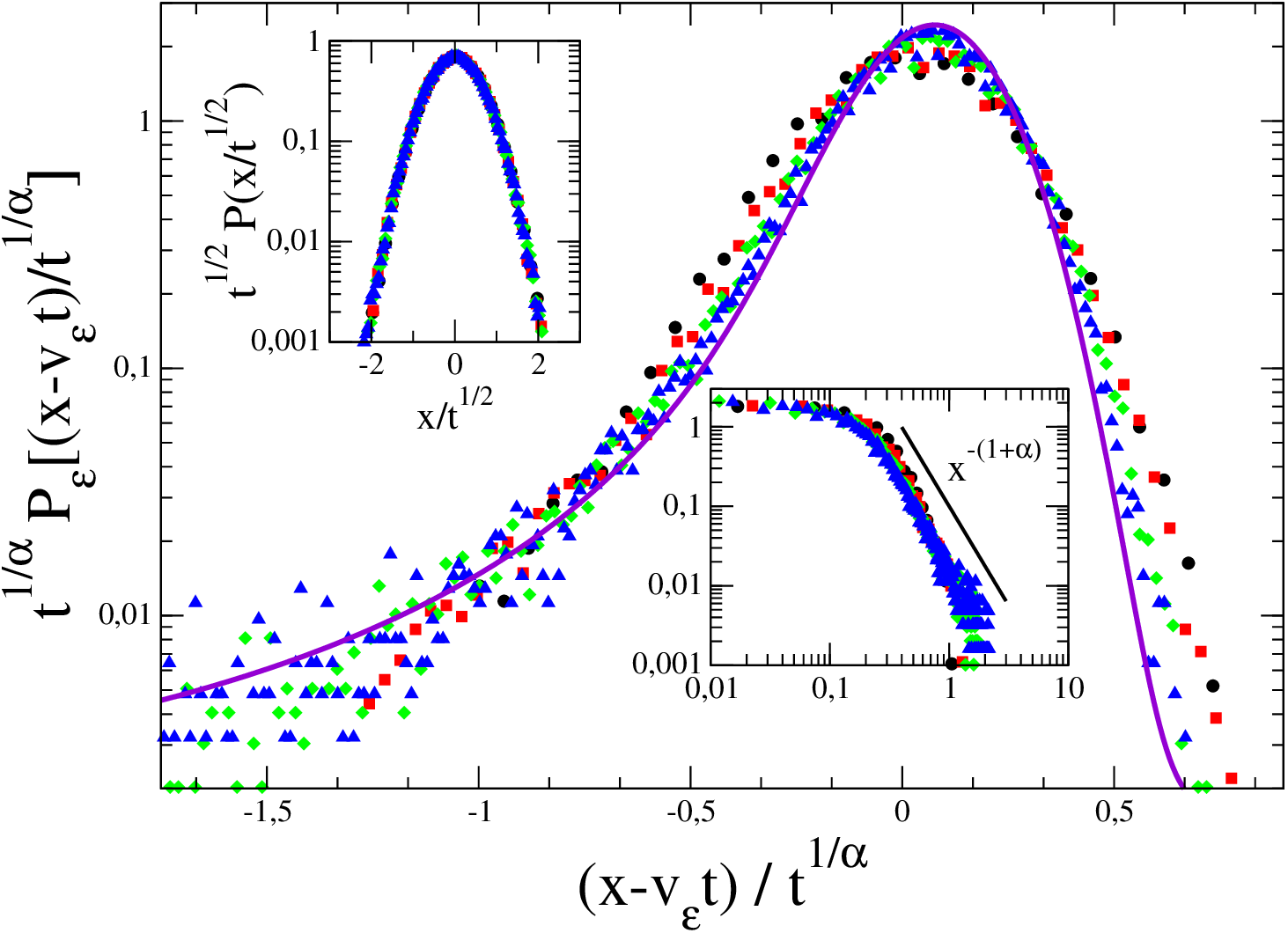}
\caption{(color online) Collapse of the PDFs of the CTRW with trapping, for
  $\alpha=1.5$ and field $\epsilon=0.15$ according to
  Eq.~(\ref{scaling}) with velocity $v_\epsilon=0.1$. Symbols
  correspond to different times: $t=10^3$ ($\bullet$), $t=2\cdot 10^3$
  ({\color{red}$\blacksquare$}), $t=5\cdot 10^3$
  ({\color{green}$\blacklozenge$}), and $t=10^4$
  ({\color{blue}$\blacktriangle$}). The continuous line represents the
  numerical inverse Fourier transform of Eq.~(\ref{four_tr}).  Central
  inset: same data of the main figure in log-log scale, as function of
  $|x-v_\epsilon t|$. Notice the power law behavior
  $P_\epsilon(x,t)\sim x^{-(1+\alpha)}$.  Top left inset: collapse of
  the PDF for $\alpha=1.5$ with zero drift, $\epsilon=0$. Notice the
  simple Gaussian behavior, $\ell(t)\sim t^{1/2}$, in agreement with
  Eq.~(\ref{four_tr_ep0}).}
\label{fig:scaling-sub-alpha1.5}
\end{figure}

For $\epsilon\not=0$ we obtain in the asymptotic regime:
\begin{equation}   
\tilde P_{\epsilon}(k,\omega)\approx
\begin{cases}
\frac{C_1 \omega^{\alpha-1}}{i \epsilon \delta_0 k+ C_2 \omega^\alpha}\qquad  \mathrm{if}\qquad 0<\alpha<1 \\
\frac{\langle t \rangle}{i (\epsilon \delta_0 k+ \langle t \rangle  \omega)+ C_3 \omega^\alpha }\qquad  \mathrm{if}\qquad 1<\alpha <2\\
\frac{\langle t \rangle}{(\delta_0 k)^2/2+ i (\epsilon \delta_0 k+ \langle t \rangle  \omega)}\qquad \mathrm{if}\qquad \alpha >2.
\end{cases}
\label{four_tr}
\end{equation}
where $C_3 (\alpha) =- \frac {C_1 (\alpha-1)}{\alpha-1}$. The immaginary parts 
of the coefficients $C_1$, $C_2$ and $C_3$ 
depend on the sign of $\omega$, so that $P_\epsilon(x,t)$ is a real
function. The numerical inverse Fourier transform of $\tilde
P_\epsilon(k,\omega)$ can be easily computed and it is in very good
agreement with the results of simulations, as shown in
Fig.~\ref{fig:scaling-sub-alpha1.5} for $\alpha=1.5$.  Notice that the
space inverse Fourier transform can be performed explicitly from
Eqs.~(\ref{four_tr}), yielding a $\Theta(x)$ function in the
$P_\epsilon(x,t)$, as expected from the arguments following
Eq.~(\ref{scaling2}). In particular, for $1<\alpha<2$ we get:
\begin{equation}   
\tilde P_{\epsilon}(x,t)= \langle t \rangle \Theta(x) \int    
e^{-i \omega (x/(\epsilon \delta_0) \langle t \rangle-t)-C_3 |\omega^\alpha | x/(\epsilon \delta_0)} d\omega. 
\end{equation}.

We remark that for $|x-v_\epsilon t|\lessapprox \ell(t)$ all the regimes
follow the general scaling pictures of Eq.~(\ref{scaling}). In
particular, for $\alpha>2$, the standard response of a diffusive
system is reproduced: $F(y)$ is a Gaussian moving at constant velocity
$v_\epsilon=\epsilon \delta_0/\langle t \rangle$. For $0<\alpha<1$,
where Gaussianity is broken already at equilibrium, the characteristic
length grows as $\ell(t)\sim t^\alpha$. We notice that $v_\epsilon=0$,
and hence a rigid motion of the probability with constant velocity is
not observed. Comparing Eqs.~(\ref{four_tr_ep0}) and~(\ref{four_tr}),
a singular behavior at $\epsilon=0$ is evident: $F(y)$ changes its
shape, as soon as $\epsilon\not=0$. The scaling function becomes
asymmetric with respect to the origin $y=0$, giving rise to an overall
motion in the direction of the external field. For $1/2 < \alpha <1$,
one finds a superdiffusive behavior induced by the field, as discussed
in~\cite{bouchaud}. In this case, due to the fast asymptotic decay of
$F(y)$, the behavior of moments is $\langle x^n(t) \rangle_\epsilon
\sim \ell^n(t)$, which is an example of \emph{weak} anomalous
diffusion~\cite{castiglione}.

The most intriguing case is for $1<\alpha<2$, where the presence of
the field turns diffusion from standard into \emph{strong} anomalous~\cite{shle}:
the scaling function moves at constant velocity $v_\epsilon=\epsilon
\delta_0/\langle t \rangle$, however the shape of $F(y)$ is not
Gaussian, and it develops a power-law tail.  In
Fig.~\ref{fig:scaling-sub-alpha1.5} we show the collapse of the PDFs
at different times, according to the scaling in Eq.~(\ref{scaling}),
for the case $\alpha=1.5$ with and without external field (see
inset). The numerical inverse Fourier transform of Eq.~(\ref{four_tr})
shows that the analytical calculations of the asymptotic regime are
able to capture the shape of the distribution and the scaling length
$l(t)\sim t^{1/z}$, with $z=\alpha<2$.  The superdiffusive spreading
of the PDF is triggered by rare events induced by long waiting times.
In particular, the exponent of the left power-law tail of the
$P_\epsilon(x,t)$ is obtained from the probability, that, up to a time
$t$, a particle remained at rest for an interval $\tau$ such that $
v_\epsilon \tau = \xi$, namely $P_\epsilon(\xi=|x-v_\epsilon t|
,t)\sim t \xi^{-(1+\alpha)} \Theta(v t - \xi)$.  Notice that the
power-law tail is suppressed for $x<0$ by the function $\Theta(x)$,
because the probability of finding particles with a negative
displacement is zero, due to the positive field. Following the above
arguments one obtains for the MSD around the peak of the distribution:
$\langle \xi^2(t) \rangle_\epsilon \sim t^{3-\alpha} \neq \ell^2(t)
\sim t^{2/\alpha}$.

\section{Solution of the master equation for the L\'evy walk}
 
Let us now consider the L\'evy walk. The master equation is:
\begin{equation}   
Q_\epsilon(x,t)=\int_0^t \frac{1}{2} \left [ Q_\epsilon(x-v_0 t'-\epsilon t'^2,t-t') 
+ Q_\epsilon(x+v_0 t'-\epsilon t'^2,t-t')\right] p_\alpha(t') dt' 
+\delta(t)\delta(x)
\label{masterlevy1}
\end{equation}
with
\begin{equation}   
P_\epsilon(x,t)=\int_0^t  \frac{1}{2} \left [Q_\epsilon(x-v_0 t'-\epsilon t'^2,t-t') 
+  Q_\epsilon(x+v_0 t'-\epsilon t'^2,t-t')\right] \int_{t'}^\infty p_\alpha(\tau)  d\tau dt', 
\label{masterlevy2}
\end{equation} 
where $p_\alpha(\tau)$ is now the distribution of flight times.
Analogously to Eq.~(\ref{me1}), we first solve Eq.~(\ref{masterlevy1}) in Fourier space and then we insert
the result in Eq.~(\ref{masterlevy2}), taking the limit of small $k$ and $\omega$.  For
$\epsilon=0$, the Fourier transform $\tilde P_0(k,\omega)$ reads~\cite{levyzero,schmi}:
\begin{equation}   
\tilde P_0(k,\omega)=
\begin{cases}
\frac{\omega^{\alpha-1} f(k/\omega)}{C_4 v_0^\alpha k^\alpha + C_2 \omega^\alpha}\qquad \mathrm{if}\qquad 0<\alpha<1 \\
\frac{\langle t \rangle}{C_4 v_0^\alpha k^\alpha + i \langle t \rangle  \omega}\qquad \mathrm{if}\qquad 1<\alpha <2\\
\frac{\langle t \rangle}{ v_0^2 \langle t^2 \rangle k^2/2 + i \langle t \rangle  \omega}\qquad  \mathrm{if}\qquad  \alpha > 2 \\
\end{cases}
\end{equation}
where $C_4$ is a real coefficient that  depends on $\alpha$, and 
$f(k/\omega)$ is a regular complex function with $f(0)=i C_1$ and 
$f(x) \sim x^{\alpha-1}$ for $x \to \infty$.
In this case, standard Gaussian diffusion is recovered for $\alpha>2$,
while for $1<\alpha<2$ and $0<\alpha<1$ the non-Gaussian propagator
corresponds to a superdiffusive ($\ell(t)\sim t^{1/\alpha}$) and a
ballistic ($\ell(t)\sim t$) motion, respectively.

\begin{figure}[!t]
\includegraphics[width=.7\columnwidth,clip=true]{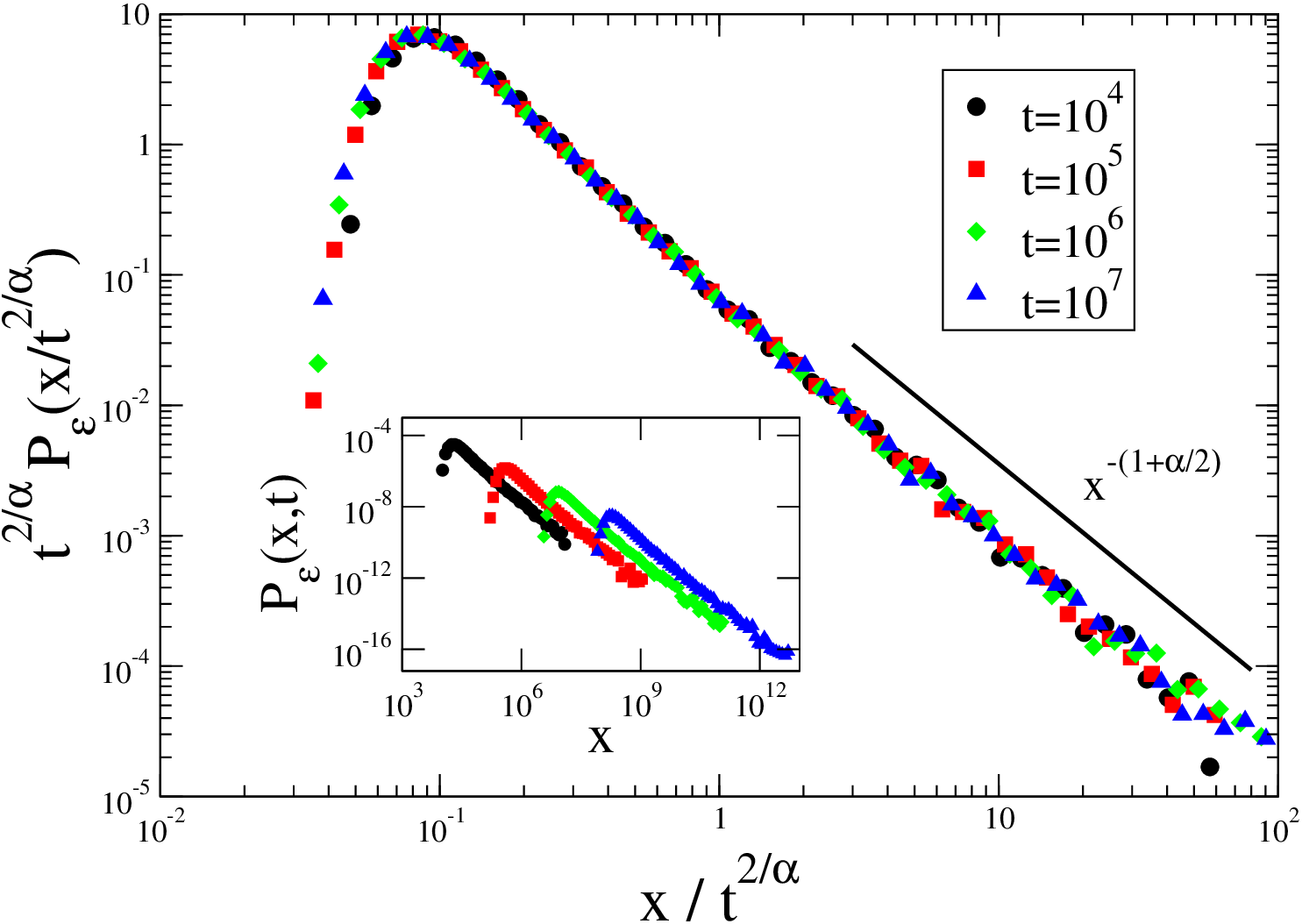}
\caption{(color online) Collapse of PDFs with the scaling $t^{2/\alpha}
  P_\epsilon(x/t^{2/\alpha})$, for the L\'evy walk with exponent
  $\alpha=1.5$ and field $\epsilon=0.25$.  Inset: $P_\epsilon(x,t)$ at
  different times. Notice that the approximation
    $(x-v_\epsilon t)/t^{2/\alpha}\sim x/t^{2/\alpha}$ works
    very well asymptotically.}
\label{fig:tails-levy-alpha1.5}
\end{figure}

\begin{figure}[!t]
\includegraphics[width=.7\columnwidth,clip=true]{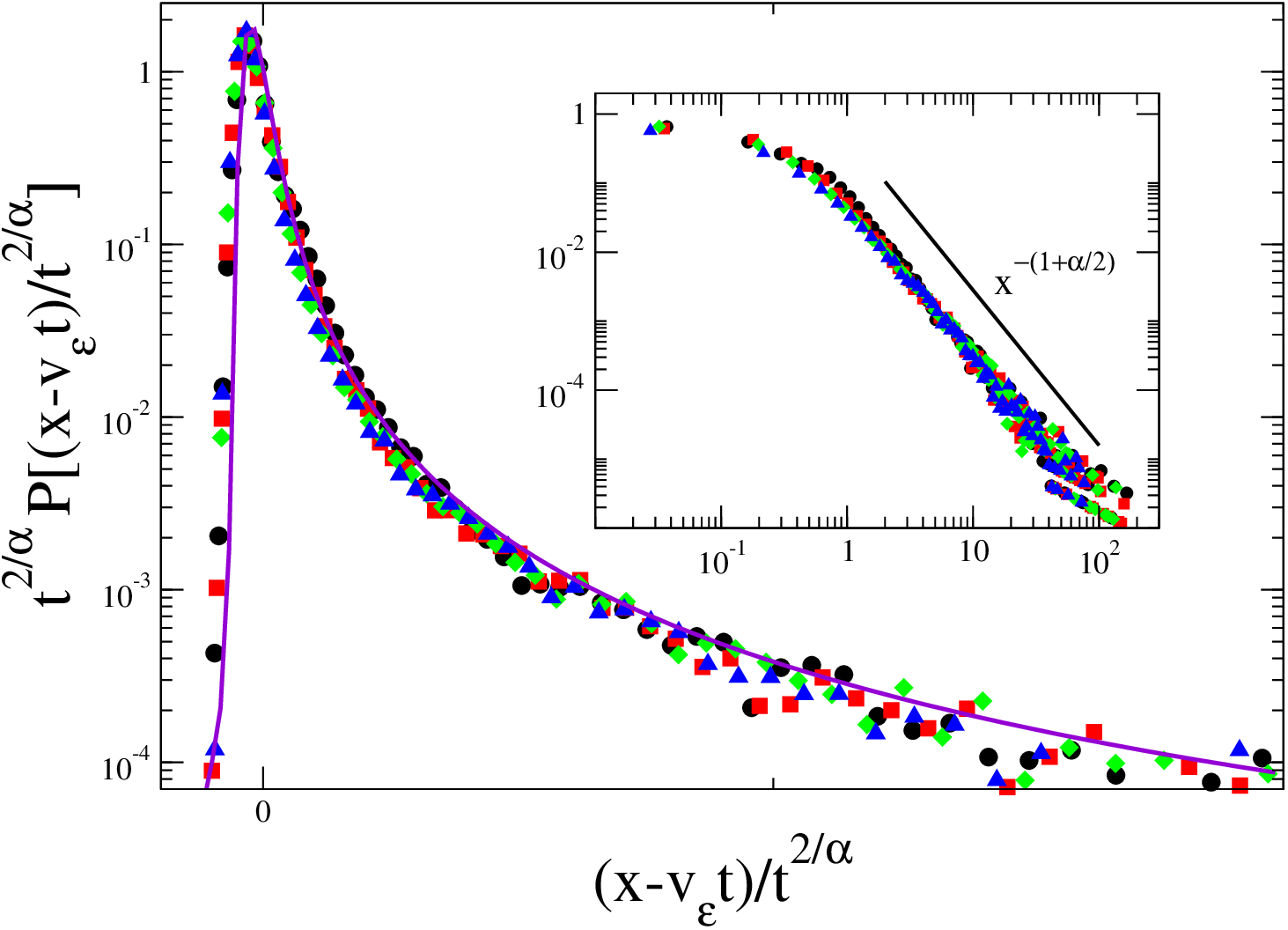}
\caption{(color online) Collapse of the PDFs with the scaling
  $t^{2/\alpha} P_\epsilon[(x-v_\epsilon t)/t^{2/\alpha}]$, for the
  L\'evy walk with exponent $\alpha=2.5$ and field $\epsilon=0.25$,
  with $v_\epsilon=0.38$. Symbols correspond to different times:
  $t=1.6\cdot 10^4$ ($\bullet$), $t=3.2\cdot 10^4$
  ({\color{red}$\blacksquare$}), $t=6.4\cdot 10^4$
  ({\color{green}$\blacklozenge$}), and $t=1.28\cdot 10^5$
  ({\color{blue}$\blacktriangle$}). The line represents the numerical
  inverse Fourier transform of Eq.~(\ref{four_levy}). Inset: same data
  of the main figure in log-log scale.}
\label{fig:tails-levy-alpha2.5}
\end{figure}

For $\epsilon\ne 0$ in the asymptotic regimes of small $k$ and
$\omega$ we obtain:
\begin{equation}   
\tilde P_{\epsilon}(k,\omega)\approx
\begin{cases}
\frac{ \omega^{\alpha-1} g(k^{1/2}/\omega) }{C_5 \epsilon^{\alpha/2} k^{\alpha/2}+ C_2 \omega^\alpha}\qquad  \mathrm{if}\qquad 0<\alpha<1 \\
\frac{ \langle t \rangle}{C_5 \epsilon^{\alpha/2} k^{\alpha/2}+ i \langle t \rangle  \omega}\qquad  \mathrm{if}\qquad 1<\alpha <2\\
\frac{ \langle t \rangle}{C_5 \epsilon^{\alpha/2} k^{\alpha/2}+ i (\epsilon \langle t^2 \rangle k+ \langle t \rangle  \omega)} \qquad \mathrm{if}\qquad  2<\alpha <4 \\
\frac{ \langle t \rangle}{  \langle t^2 \rangle v_0^2 k^2/2+ i (\epsilon \langle t^2 \rangle k+ \langle t \rangle  \omega)} \qquad \mathrm{if}\qquad \alpha >4
\end{cases}
\label{four_levy}
\end{equation}
where $C_5(\alpha)$ is a complex number whose immaginary part
depends on the sign of $\omega$ and 
$g(x)$ is a regular complex function with $g(0)=i C_5$ and 
$g(x) \sim x^{\alpha-1}$ for $x \to \infty$.

Retaining only the leading terms in $k$ and $\omega$ in the previous
expressions and performing the inverse Fourier transform in space and time,
one obtains a PDF of displacements in very good agreement with the
numerical simulations (see
Fig.~\ref{fig:tails-levy-alpha2.5}). However, the asymptotic
approximation looses the cut-off on displacements at a finite time.
Indeed, at each time $t$, the largest allowed displacement is
$x=v\tau+\epsilon \tau^2\sim \epsilon \tau^2$, corresponding to the
rare event of a particle never colliding up to $\tau=t$, and this
bound necessarily requires for the $\Theta(\epsilon t^2 - x)$ in
Eq.~(\ref{scaling2}).

We find that for $\alpha>4$ Eq.~(\ref{four_levy}) describes a Gaussian
distribution, moving at velocity $v_\epsilon=\epsilon \langle t^2
\rangle/\langle t \rangle$. For $\alpha<2$, the rigid translation of
the scaling function is subdominant ($v_\epsilon t\ll t^{2/\alpha}$)
and the global effect of the field is a strong asymmetry of $F(y)$
(see Fig.~\ref{fig:tails-levy-alpha1.5}).  In this case the scaling
length grows in super-ballistic way, i.e. $z=1/2$ (linearly
accelerated motion) for $0<\alpha<1$ and $z=\alpha/2$ for
$1<\alpha<2$. Finally, for $2<\alpha<4$ the drift with velocity
$v_\epsilon$ becomes relevant, but the scaling function is still not
Gaussian due to an emerging power-law tail which produces a strong
anomalous superdiffusive behavior.  The remarkable feature for the
L\'evy walk with $2<\alpha<4$ is that, similarly to the CTRW
with $1<\alpha<2$, the field $\epsilon$ produces a power-law tail
otherwise not present in the zero-field Gaussian distribution (see
Fig.~\ref{fig:tails-levy-alpha2.5}). Moreover the scaling length
$\ell(t)$ which yields the collapse of the PDFs does not govern the
behavior of moments, $\langle x^n(t) \rangle_\epsilon \neq \ell^n(t)$.
The algebraic tail of $P_\epsilon(x,t)$ at large time can be obtained
from the distribution of flight times by changing variables: one finds
$P_\epsilon(x,t) \sim x^{-(1+\alpha/2)}$, yielding the asymptotic
behavior readable in Figs.~\ref{fig:tails-levy-alpha1.5}
and~\ref{fig:tails-levy-alpha2.5}. This power-law behavior, together
with the cut-off $\Theta(\epsilon t^2 - x)$, yields the MSD around the
peak of the distribution: $\langle [\delta
  x(t)]^2\rangle_\epsilon=\langle x^2(t)\rangle_\epsilon-\langle
x(t)\rangle^2_\epsilon \sim t^{5-\alpha} \neq \ell(t)^2\sim
t^{4/\alpha}$, showing that a single scaling length cannot capture the
behavior of all moments.

\section{Conclusion}

We have presented results for the CTRW with trapping and the L\'evy walk  in
the presence of currents. At variance with standard Gaussian systems,
the field, even arbitrarily small, significantly modifies the scaling
properties of the PDF of displacements, introducing a new length-scale
related to rare events.  Beyond the principal scaling length of the
distribution $\ell(t)$ and that related to the rigid shift,
$\ell_T(t)$, one must also consider the typical length introduced by
the cut-off of the power-law tail, $\ell_e(t)$, necessary for the
calculation of higher order moments. This is how rare events induce
the \emph{strong} anomalous behavior.  The change in transport
properties in the presence of an external field represents a valuable
probe to unveil the underlying dynamical structure of the system.
Field-induced anomalous behavior highlights the importance of rare and
large fluctuations in regimes where the diffusional properties are
\emph{apparently} standard. 

Our results have been derived in a one-dimensional model, both for the
CTRW with trapping and the L\'evy walk. However, we expect that the
main effects should be valid also in realistic experimental three
dimensional samples, due to the decoupling of the motion in the
direction of the field and in the orthogonal direction. Hence, we
expect to observe the same behavior along the applied field, while the
motion in the perpendicular directions should be described by the CTRW
equations without fields i.e. $P_0(x,t)$.

\ack
We thank A. Puglisi and D. Villamaina for useful discussions. RB, AS and AV thank the Kavli Institute for Theoretical Physics China at the Chinese Academy of Sciences, Beijing, for the kind hospitality
at the workshop "Small system nonequilibrium fluctuations, dynamics and stochastics, and anomalous behavior" July 2013 where the paper was completed. The work of GG and AS is supported by the Granular Chaos project, funded by the Italian MIUR under the grant number RIBD08Z9JE.


\section*{References}

\begin{thebibliography}{99}

\bibitem{bardou}
F.~Bardou, J.-P.~Bouchaud, A.~Aspect, and C.~Cohen-Tannoudji, 2001
\emph{L\'evy Statistics and Laser Cooling: How Rare Events Bring Atoms to Rest}
(Cambridge University Press, Cambridge).

\bibitem{wang}
B.~Wang, J.~Kuo, S.~C.~Bae, and S.~Granick, 2012
\emph{Nature Materials}  {\bf 485} 11 .

\bibitem{tang}
L.~H.~Tang and H. Chat\'e, 2001
\emph{Phys. Rev. Lett.}  {\bf 86} 830. 

\bibitem{sanchez}
A.~D.~S\'anchez, J.~M.~L\'opez,  M.~A.~Rodr\'iguez, and M.~A.~Mat\'ias, 2004
\emph{Phys. Rev. Lett.}  {\bf 92} 204101. 

\bibitem{bouchaud}
J.-P.~Bouchaud and A.~Georges, 1990
\emph{Phys. Rep.}  {\bf 195} 127.

\bibitem{ribiere}
P.~Ribi\`ere, P.~Richard, R.~Delannay, D.~Bideau, M.~Toiya, and W.~Losert, 2005 
\emph{Phys. Rev. Lett.} {\bf 95} 268001.

\bibitem{beni1}
O.~B\'enichou, C.~Mej\'ia-Monasterio, and G. Oshanin, 2013
\emph{Phys. Rev. E}  {\bf 87} 020103(R). 

\bibitem{beni3}
O.~B\'enichou, P.~Illien, C.~Mejía-Monasterio, and G.~Oshanin, 2013
\emph{J. Stat. Mech.}  P05008.

\bibitem{wilson}
L.~G.~Wilson and W.~C.~K.~Poon, 2011
\emph{Phys. Chem. Chem. Phys.} {\bf 13} 10617.

\bibitem{barkai}
E.~Barkai and V.~Fleurov,  1998
\emph{Phys. Rev. E} {\bf 58} 1296. 

\bibitem{jespersen}
S.~Jespersen, R.~Metzler, and H.~C.~Fogedby, 1999
\emph{Phys. Rev. E}  {\bf 59} 2736.

\bibitem{villamaina}
D.~Villamaina, A.~Sarracino, G.~Gradenigo, A.~Puglisi and A.~Vulpiani, 2011
\emph{J. Stat. Mech.}  L01002. 

\bibitem{bettolo}
U.~Marini Bettolo Marconi, A.~Puglisi, L.~Rondoni, and A.~Vulpiani, 2008
\emph{Phys. Rep.}  {\bf 461} 111.

\bibitem{beni2}
O.~B\'enichou, P.~Illien, G.~Oshanin, and R.~Voituriez, 2013
\emph{Phys. Rev. E} {\bf 87} 032164.

\bibitem{angelani}
L.~Angelani, R.~Di Leonardo, G.~Parisi, and G.~Ruocco,  2001
\emph{Phys. Rev. Lett.} {\bf 87} 055502.

\bibitem{levitz}
P.~Levitz,   593 
\emph{Europhys. Lett.} {\bf 39} (1997).

\bibitem{havlin}
D.~ben-Avraham and S.~Havlin, 2004
\emph{Diffusion and Reactions in Fractals and Disordered Systems (Cambridge University Press}
(Cambridge University Press).

\bibitem{Levyrandom}
R.~Burioni, L.~Caniparoli, and A.~Vezzani, 2012
\emph{Phys. Rev. E} {\bf 81} 060101.

\bibitem{Levy2d}
P.~Buonsante, R.~Burioni and A.~Vezzani, 2011
\emph{Phys. Rev. E}  {\bf 84} 021105.

\bibitem{benichou}
O. B\'enichou, C. Loverdo, M. Moreau,  and R. Voituriez,  2011
\emph{Rev. Mod. Phys.} {\bf 83} 81.

\bibitem{shles1985}
M.~F.~Shlesinger and J.~Klafter,  1985
\emph{Phys. Rev. Lett.} {\bf 54} 2551.
   
\bibitem{barthelemy}
P.~Barthelemy, J.~Bertolotti and D.~S.~Wiersma,  2008
\emph{Nature} {\bf 453} 495.

\bibitem{bertolotti} 
J. Bertolotti, K. Vynck, L. Pattelli, P.~Barthelemy, S.~Lepri, and D.~S.~Wiersma,  2012 
\emph{Adv. Funct. Mat.}  {\bf 20} 965. 

\bibitem{klages}
R.~Klages, G.~Radons and I.~M.~Sokolov (Eds.), 2008
\emph{Anomalous Transport: Foundations and Applications}
(Wiley, VCH Berlin).

\bibitem{lechenault}
F.~Lechenault, R.~Candelier, O.~Dauchot, J.-P.~Bouchaud and G.~Biroli, 2012 
\emph{Soft Matter} {\bf 6} 3059. 

\bibitem{winter}
D.~Winter, J.~Horbach, P.~Virnau and K.~Binder, 2012 
\emph{Phys. Rev. Lett.}{\bf 108} 028303. 

\bibitem{heuer}
C.~F.~E.~Schroer and A.~Heuer, 2013 
\emph{Phys. Rev. Lett.} {\bf 110} 067801.

\bibitem{castiglione} 
P.~Castiglione, A.~Mazzino, P.~Muratore-Ginanneschi, and A.~Vulpiani,  1999
\emph{Physica D} {\bf 134} 75.

\bibitem{shle}
M. Shlesinger,   1974
\emph{J. Stat. Phys.} {\bf 10} 421 

\bibitem{levyzero}
M.~Zumofen and J.~Klafter, 1993
\emph{Phys. Rev. E}  {\bf 47} 851.

\bibitem{schmi}
M.~Schmiedeberg, V.~Y.~Zaburdaev, and H.~Stark, 2009
\emph{J. Stat. Mech.} P12020.

\bibitem{sokolov}
I.~M.~Sokolov and R.~Metzler,  2003
\emph{Phys. Rev. E} {\bf 67} 010101(R).

\bibitem{gradenigo}
G.~Gradenigo, A.~Sarracino, D.~Villamaina, and A.~Vulpiani,  2012
\emph{J. Stat. Mech.} L06001.

\bibitem{klafter}
J.~Klafter and I.~M.~Sokolov, 2011
\emph{First steps in random walks}
(Oxford University Press, Oxford).

\end{thebibliography}
\end{document}